\documentclass[prb,aps,amsmath,amssymb,amsfonts,floatfix,superscriptaddress,showpacs,twocolumn]{revtex4}
\usepackage{graphicx}
\usepackage[dvips]{color}
\usepackage{dcolumn}

\newcommand{\bk}{\mathbf{k}}

\newcommand{\br}{\mathbf{r}}

\newcommand{\fig}[1]{Fig.~\ref{#1}}

\newcommand{\mub}[0]{$\mu_{\rm B}$}

\def\bra{\langle}
\def\ket{\rangle}

\begin{document}

\title{Importance of electronic correlations for structural and
  magnetic properties of the iron pnictide superconductor LaFeAsO} 

\author{Markus~Aichhorn}
\affiliation{Institute of Theoretical and Computational Physics, TU Graz, Petersgasse 16, Graz, Austria}
\affiliation{Centre de Physique Th\'eorique, \'Ecole Polytechnique, CNRS,
91128 Palaiseau Cedex, France}
\author{Leonid~Pourovskii}
\affiliation{Centre de Physique Th\'eorique, \'Ecole Polytechnique, CNRS,
91128 Palaiseau Cedex, France}
\affiliation{Division of Theory and Modeling (IFM), Link\"oping University, SeRC, SE-581 83 Link\"oping, Sweden}
\author{Antoine~Georges}
\affiliation{Centre de Physique Th\'eorique, \'Ecole Polytechnique, CNRS,
91128 Palaiseau Cedex, France}
\affiliation{Coll\`ege de France, 11 place Marcelin Berthelot, 75005
  Paris, France}
\affiliation{DPMC, Universit\'e de Gen\`eve, 24 quai E. Ansermet, CH-1211 Gen\`eve, Suisse}

\begin{abstract}
We present calculations of structural and magnetic properties of the
iron-pnictide superconductor LaFeAsO including electron-electron
correlations. For this purpose we apply a fully charge self-consistent
combination of Density-Functional Theory with the Dynamical Mean-Field
theory, allowing for the calculation of total energies. We find that
the inclusion of correlation effects gives a good agreement of
the Arsenic $z$ position with experimental data even in the
paramagnetic (high-temperature) phase. Going to low temperatures, we
study the formation of the ordered moment in the striped
spin-density-wave phase, yielding an ordered moment of about $0.60$\,\mub, again in
good agreement with experiments. This shows that the inclusion of
correlation effects improves both structural and magnetic properties
of LaFeAsO at the same time.
\end{abstract}

\pacs{71.15.Mb, 71.20.Be, 74.70.-b}
\maketitle

\section{Introduction}\label{sect:intro}

Since the discovery of high-temperature superconductivity in
iron-based compounds\cite{kamihara1} a lot of research 
has been dedicated to this fascinating class of materials. On the theoretical
side, calculations within density-functional theory (DFT), often performed within the 
local density approximation (LDA), could reproduce a 
variety of properties, but failed in the quantitative description of other 
features, such as the mass renormalization of the predominately iron-$d$ 
quasi-particles, which could be improved by the inclusion of
correlation effects for the Fe 3d electrons.\cite{haule1,haule2,anisimov2,aichhorn2009}
A very puzzling mystery shows up for the compounds exhibiting 
long-range spin-density-wave (SDW) magnetic ordering at low temperatures. 
Although the spin-pattern and ordering vectors were well predicted by DFT, 
it was soon realized that there is a big discrepancy between the magnitude 
of the measured magnetic moments with theoretical predictions of spin-polarized 
DFT calculations. For instance for LaFeAsO, early experimental data pointed to a very 
small ordered moment in the range of range of
around 0.3\,\mub,\cite{delacruz1,mcguire2008} although recent  
measurements indicated a somewhat larger moment of 0.63\,\mub{}
in LaFeAsO.\cite{qureshi2010} On the 
other hand, DFT calculations using the 
experimental crystal structure always gives large values between 1.7 and more than 
2\,\mub.\cite{dong1,cao1,ma2008,mazin1,ishibashi2008,yin-z-p1}

There is also a strong connection between the value of the ordered moment and 
details of the crystal structure. As stated above, the ordered moment turns 
out to be too large in DFT calculations, but in this magnetic case, structural 
optimization of the $z$-position of the arsenic ions reproduces well the 
experimental position.\cite{mazin1} On the other hand, non-magnetic DFT calculations, which 
should correspond to the paramagnetic high-temperature phase, gives a too short 
Fe-As distance with drastic influence on the low-energy electronic structure.\cite{yin-z-p1} 
Experimentally, the As $z$ position hardly changes across the
magnetic transition,\cite{nomura2008,delacruz1}  
a fact that is hard to reconcile within DFT calculations, since the
optimized internal structural parameters differ significantly between
magnetic and non-magnetic calculations. The correct 
description of the equilibrium structure is particularly important for cases where 
the forces on the ions are important, e.g. phonon calculations. 

There were several attempts to improve over simple DFT calculations. Concerning 
the ordered moment, Yildirim {\em et al.}\cite{yildrim2008_1} performed fixed
moment DFT calculations in order to study the stability of magnetic
ordering patterns. 
Attempts to include correlation effects by performing LDA+U or GGA+U
calculations were not successful. It has been
shown\cite{ishibashi2008_2} that the magnetic moment even {\em
  increases} with increasing $U$, and even for small interaction
values of $U\approx 1$\,eV the topology of the Fermi surfaces is
changed drastically, incompatible with experiments. 
However, 
a reduction of the magnetic  
moment could be found in LDA+U calculation using an 
effective 
{\em negative} interaction parameter.\cite{ferber2010} 
Particularly promising are approaches 
using many-body techniques to include electronic correlation effects. 
Using a combination of DFT with the dynamical mean-field theory (DMFT)  
a significant reduction of the magnetic moment could be found for
BaFe2As2,\cite{yin2011}
variational Monte Carlo gave similar results also for other
materials,\cite{misawa2011} consistent with a recent comprehensive
LDA+DMFT study for a variety of pnictide and chalcogenide materials.\cite{yin2011u}
A general argument is that quantum fluctuations hinder a large instantaneous iron 
moment from ordering.\cite{hansmann2010,lee_hunpyo2010,yin2011,misawa2011,zhang2010,yin2011u} 

Regarding the combination of structural  
and magnetic properties, one proposal for a better description is to combine in a 
sophisticated  way magnetic and non-magnetic DFT calculations.\cite{boeri2010} This approach 
has been used to study the electron-phonon interaction in iron-pnictide superconductors. 
Improved structural optimization has been performed using a combination of DFT 
with Gutzwiller wave function techniques,\cite{wang2010} where the values of the 
interaction parameters where fitted to give the correct As height
above the Fe plane.

The motivation for this article is to show that the inclusion of correlation effects 
by LDA+DMFT for the description of LaFeAsO improves substantially the agreement of both 
the As $z$ position as well as the ordered magnetic moment between theory and experiment 
within one set of {\em ab-initio} calculated interaction parameters, which are 
determined with the constrained Random Phase Approximation
(cRPA).\cite{miyake2010,miyake1} 
This kind of investigation has not been done so far, since theoretical
studies including strong electron-correlations were focused on the
calculation of {\em either}
magnetic\cite{hansmann2010,yin2011,yin2011u} {\em or} structural
properties.\cite{wang2010} 

A consistent approach to total energy calculations and structural
optimization within LDA+DMFT (the  As $z$ position in LaFeAsO in the
present work) requires self-consistency over the charge
density. LDA+DMFT is often employed within the so-called "one-shot"
scheme, where the one-electron part of the Hamiltonian obtained from
the band-structure LDA part is not updated during subsequent DMFT
calculations. However, correlation effects will in general induce a
certain redistribution of the charge density, which in turn leads to
different Kohn-Sham potential and one-particle part of the
Hamiltonian. The correlation-induced changes in the charge density and
one-electron potential will also affect the electron-nuclei, Hartree,
and exchange-correlation contributions to the LDA+DMFT total
energy. Moreover, in some systems (e.g. cerium oxides \cite{pour2007})
the charge density self-consistency has been demonstrated to be
important for spectral properties as well. 

The paper is organized as follows. In Sect.~\ref{sect:method} we
introduce the full charge-self consistent LDA+DMFT method, followed by
Sect.~\ref{sect:results} where we present results for the LaFeAsO
system. We draw our conclusions in Sect.~\ref{sect:conclusions}, which
are followed by App.~\ref{sect:appendix} with a more detailed
discussion of the influence of the full charge self-consistency on the
single-particle spectra.

\section{Methods}\label{sect:method}

For the present study, we use a further development of a previously introduced LDA+DMFT 
implementation, Ref.~\onlinecite{aichhorn2009}, which is based on the
full potential (linearized) augmented plane wave method as implemented
in the Wien2K package.\cite{Wien2K} Our task of optimizing
the arsenic ion position necessitates rather accurate calculations of
the total energy, which, as explained in the Sec.~\ref{sect:intro},
require a  LDA+DMFT scheme fully self-consistent in the charge
density. The implementation of full charge self-consistency is
currently a topic of high interest, and several schemes have
been implemented recently.\cite{savrasov2004,pour2007,haule2009,amadon2011u}

Within the projective technique for formation of the correlated
orbitals, Ref.~\onlinecite{aichhorn2009}, we use the Kohn-Sham (KS)
states within a chosen energy window $\mathcal{W}$ to form
Wannier-like functions that are treated as correlated
orbitals. 
In the present work, we use an energy window from $-6.8$\,eV to
  $2.8$\,eV, spanning the range of Fe-d as well as As-p and O-p
  states, giving a total number of 22 bands inside the
  window. Onsite interactions were then applied to the five Fe-d
  orbitals. The very same projection scheme has already been used in
  Ref.~\onlinecite{aichhorn2009}. 
Solving the corresponding 
single-site
quantum impurity problem produces 
the local self-energy within the correlated orbitals basis set, which
is then upfolded into the lattice self-energy
$\Sigma_{\nu\nu'}(\mathbf{k},i\omega_{n})$, where $\omega_{n}$ are
Matsubara fermionic frequencies . The lattice self-energy
$\hat{\Sigma}(\mathbf{k},i\omega_{n})$ is generally non-diagonal in
the subspace of the KS eigenstates $\{\nu\}$ ($\nu \in \mathcal{W}$)
leading to a non-diagonal lattice Green's function within
$\mathcal{W}$ and to the corresponding density matrix: 

\begin{equation}\label{N_nunu}
N^{k}_{\nu\nu'}=\sum_{n} G_{\nu\nu'}(\mathbf{k},i\omega_n)e^{i\omega_n 0^+}
\end{equation}
being also non-diagonal. The charge density distribution in the real
space is then calculated from the density matrix $N^{k}_{\nu\nu'}$ as
follows: 

\begin{equation}\label{rho_general_dmft}
\rho_{DMFT}(\mathbf{r})=\rho_{ow}(\mathbf{r})+\sum_{k,\nu\nu'}\langle
\mathbf{r}|\Psi_{k\nu}\rangle N^{k}_{\nu\nu'} \langle
\Psi_{k\nu'}|\mathbf{r} \rangle , 
\end{equation}
where $\Psi_{k\nu}$ are the KS eigenstates within the  energy window
$\mathcal{W}$, $\rho_{ow}(\mathbf{r})$ is the contribution from states
outside  $\mathcal{W}$. By substituting into (\ref{rho_general_dmft})
the expansion of the KS eigenstates within the linear augmented
plain-wave (LAPW) basis set one derives formulas for the charge
density within the muffin-tin (MT) spheres and in the
interstitial. These formulate are generalizations of the standard LAPW
expressions to the case of a density matrix non-diagonal in the space
of KS states. As in the standard case, the charge density within the
MT spheres is expressed through radial solutions (and their energy
derivatives) of the corresponding Schr\"{o}dinger equation. In the
interstitial it is expressed through plain waves. The derivation and
relevant formulas for each case are given in Appendix
\ref{sect:appendix_chd}. 

The LDA+DMFT total energy reads\cite{amadon2006}
\begin{equation}\label{E_tot}
\begin{split}
E= & E_{kin} + E_{c}[\rho_{DMFT}]+E_H[\rho_{DMFT}]+ \\
 & E_{xc}[\rho_{DMFT}]  + \bra H_U \ket - E_{DC}, 
\end{split}
\end{equation}
where the corresponding contributions in the right-hand side are the
kinetic, crystal (electron-nuclei and nuclei-nuclei),  Hartree,
exchange-correlation, Hubbard and double-counting correction terms,
respectively. The second, third and four terms are evaluated in
accordance to the standard DFT-LDA expressions but with the updated
LDA+DMFT charge density (\ref{rho_general_dmft}). 
The kinetic energy contribution reads 
\begin{equation}\label{e_kin}
E_{kin}= E_{band}-\int d\br v_{KS}(\br) \rho_{DMFT}(\br) ,
\end{equation}
where the Kohn-Sham potential $v_{KS}$ corresponds to the LDA+DMFT charge density $\rho_{DMFT}$
and the band energy contribution $E_{band}$ is
\begin{equation}\label{e_band}
E_{band}=E_{band}^{ow}+\sum_{\bk} {\rm Tr} \hat{H}^k_{KS}\hat{N}^{k} = E_{band}^{ow}+ \sum_{\bk \nu}\epsilon_{k\nu} N^{k}_{\nu\nu},
\end{equation}
 where ${H}^k_{KS}$ is the one-particle (Kohn-Sham) part of the
Hamiltonian, $\epsilon_{k\nu}$ are its eigenstates with $\nu \in
\mathcal{W}$, and $E_{band}^{ow}$ is the sum over the occupied
Kohn-Sham eigenstates laying outside 
of the window $\mathcal{W}$. 

Finally, the Hubbard term $\bra H_U \ket$ was evaluated in accordance
with the Migdal formula $\bra H_U \ket = \frac{1}{2}{\rm
  Tr}\,({\mathbf\Sigma}(i\omega) {\mathbf G}(i\omega))$, where
${\mathbf\Sigma}(i\omega)$ and ${\mathbf G}(i\omega))$ are the
impurity self-energy and Green function, respectively.

For the solution of the quantum impurity problem we apply the
continuous-time quantum Monte Carlo method in the strong-coupling
formulation.\cite{werner_ctqmc} Restricting ourselves to
density-density interactions only, as in
Ref.~\onlinecite{aichhorn2009}, we are able to perform calculations
down to temperatures as low as $T=77$\,K with reasonable numerical
effort and without further approximations. For the calculation of
total energies, high-quality numerical 
data is necessary. In order to get an estimate of the statistical
error on the total energy, we perform several further iterations
(order 10) after self-consistency is reached, yielding an estimate for
the standard deviation.

Interaction parameters have been calculated previously within
cRPA\cite{aichhorn2009}, yielding an average Coulomb interaction of
$U=2.7$\,eV and Hund's exchange of $J=0.8$\,eV. For details of this
calculation we refer the reader to
Refs.~\onlinecite{aichhorn2009,miyake2010,miyake1}. 

As mentioned above, a double-counting correction has to be applied in
order to subtract the contribution to the correlation energy already
included in the LDA. Several forms have been proposed, we will apply
the two most common approximations, which are the around-mean-field
(AMF) and the full-localized-limit (FLL) forms,
\begin{align}
\Sigma^{\sigma,\rm AMF}_{DC}&=U\left( N-n \right)-J\left(N_\sigma -n\right)\\
\Sigma^{\sigma,\rm FLL}_{DC}&=U\left(N - 0.5\right)-J\left(N_\sigma -0.5\right),
\end{align}
where $N$ is the total electronic charge of the impurity problem,
$N_\sigma$ its spin-dependent value, and $n$ the charge per spin and
orbital. For the corresponding double counting energies one can
find\cite{ylvisaker2009} 
\begin{align}
E^{\rm AMF}_{DC}&=\frac{1}{2}UN^2-\frac{U+2lJ}{2l+1}
\frac{1}{2}\sum_\sigma N_\sigma^2\\
E^{\rm FLL}_{DC}&=\frac{1}{2}UN(N-1)-\frac{J}{2}\sum_\sigma N_\sigma(N_\sigma-1),
\end{align} 
with $l=2$ the orbital quantum number for $3d$ electrons.

Since LSDA calculations give a highly polarized state, we perform our
spin-polarized DMFT calculations starting from non-magnetic LDA
calculations.

\section{Results}\label{sect:results}

\begin{figure}[ht]
  \centering
  \includegraphics[width=0.9\columnwidth]{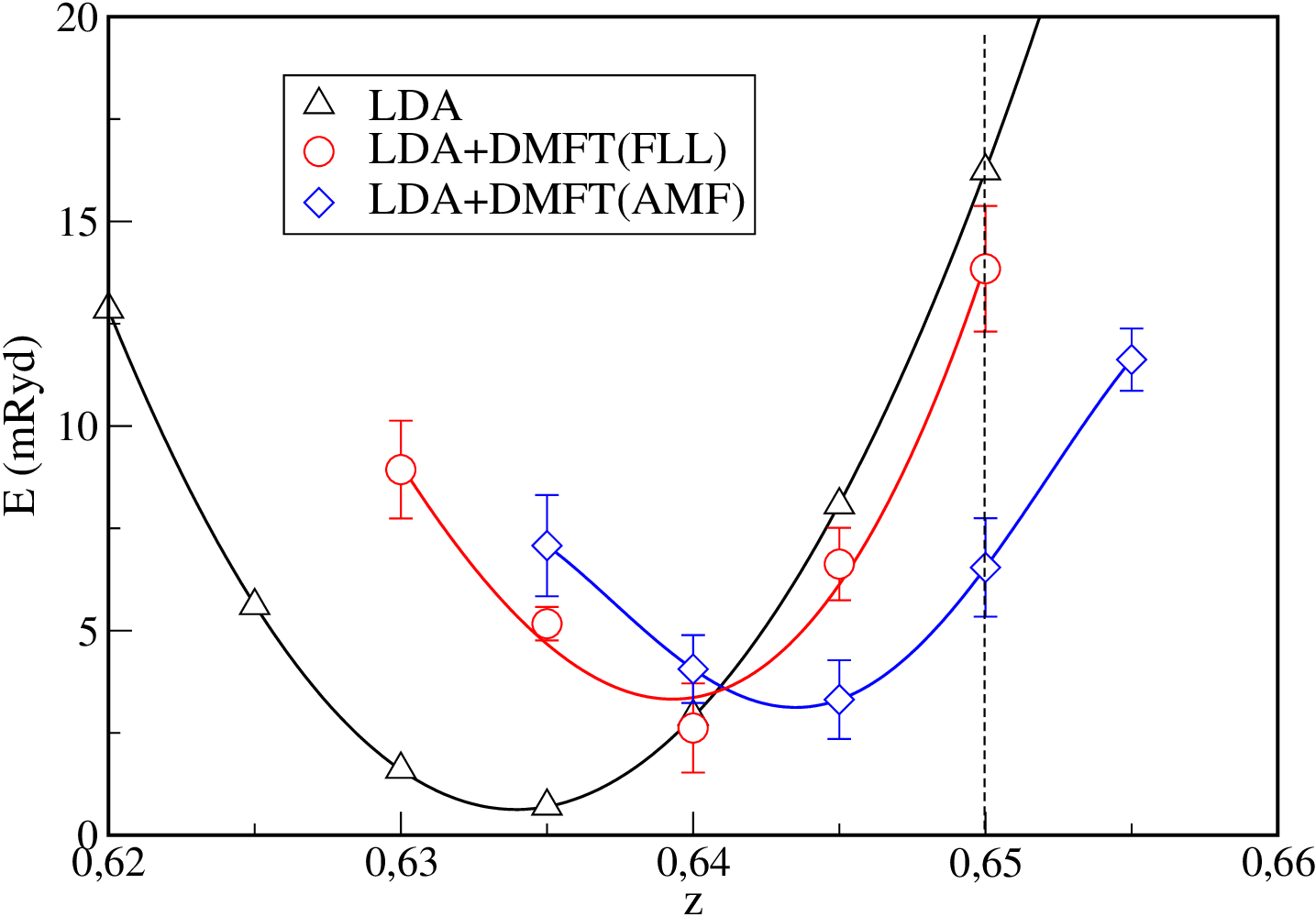}
  \caption{\label{fig:totenerg}%
    (Color online) The relative total energy of LaFeAsO as function of the As height in 
    the unit cell ($z$-parameter). Solid black line (triangles): LDA result. Red line (circles): 
    LDA+DMFT using FLL double counting. Blue line (diamonds): LDA+DMFT using AMF double counting. 
    Curves are shifted to give similar absolute value of the total energy. Vertical dashed line 
    marks the experimental $z$ position. Error bars are calculated
    from averaging several further iterations at the self-consistent solution.
  }
\end{figure}
\begin{figure}[t]
  \centering
  \includegraphics[width=0.9\columnwidth]{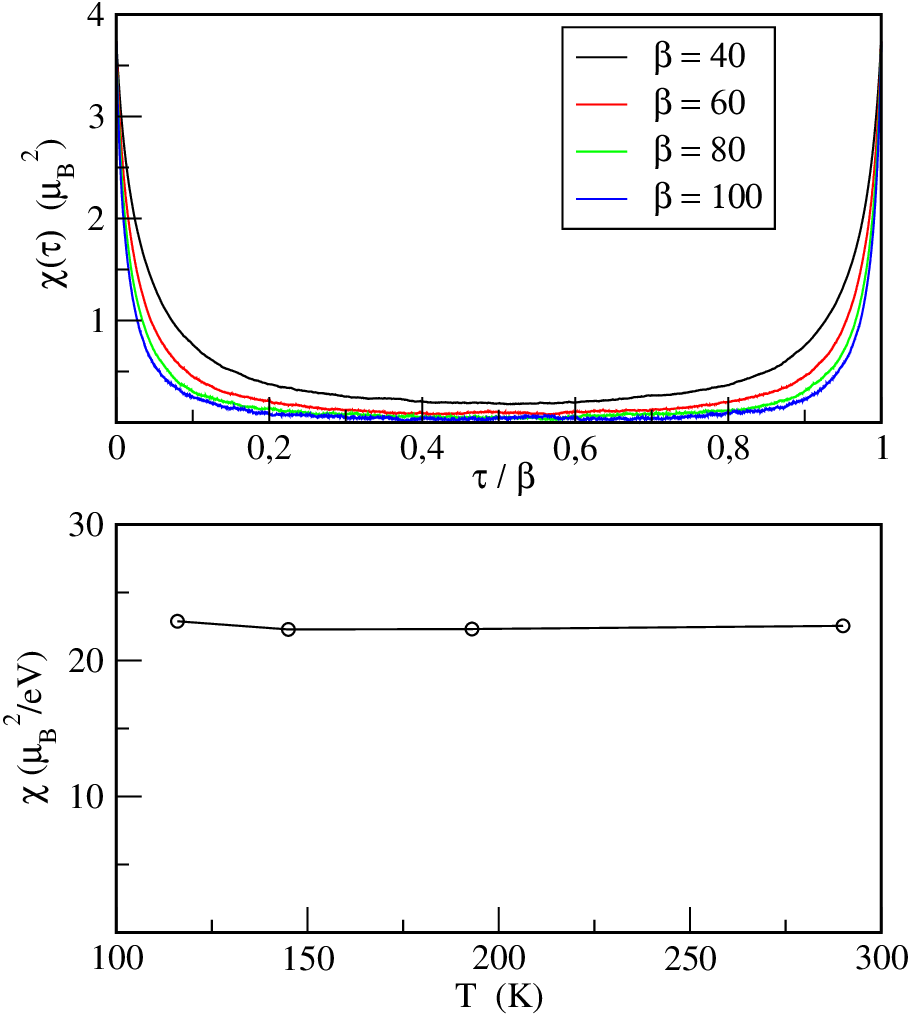}
  \caption{\label{fig:magn}%
    (Color online) Magnetic properties of LaFeAsO in the paramagnetic
    (high-temperature) phase, 
    calculated using AMF double counting corections. 
    Upper panel: Local spin susceptibility for inverse temperatures (from top to bottom) $\beta=40$, 
    $60$, $80$, and $100$\,eV$^{-1}$, corresponding to temperatures $T=290$, $193$, $145$, and $116$\,K. 
    Lower panel: Static susceptibility. 
  }
\end{figure}

Let us start the discussion of our result with the determination of the As $z$ position. We did 
paramagnetic LDA+DMFT calculations at inverse temperature $\beta=40$\,eV$^{-1}$, roughly corresponding 
to room temperature, using the two different types of double counting
corrections mentioned in 
Sect.~\ref{sect:method}. In \fig{fig:totenerg} we compare the results with the structure 
optimization within the LDA, calculated using the Wien2K package. It is obvious from these 
curves that the inclusion of correlation effects via the DMFT significantly improves over 
the LDA results. The ion is pushed away from the iron layers towards the experimentally realized 
$z$ position. Interestingly, the choice of the double-counting
correction, although having almost no effect on the  
single-particle spectra (see below, Sect~\ref{sect:appendix}), has some visible effect on 
the total energy. This is most likely due to the very small energy scales that one has to 
deal with in these structure optimizations, where already tiny differences can visibly show 
up. Nevertheless, using the AMF double counting, the As $z$ position as determined in LDA 
calculations ($z\approx0.634$) is corrected to around $z=0.643$, which has to be compared 
with the experimental value of $z=0.651$. We attribute the larger distance of the As ion 
from the iron plane to the fact that in DMFT calculations the ground state of the iron atom is 
the S=2 high spin state, having slightly larger ionic radius then the non-magnetic state realized 
in non-magnetic LDA calculations.

The small discrepancy that we still see between our calculated $z$ values and experimental 
data is most likely due to the neglection of Coulomb interactions between the iron and arsenic 
ions ($p$-$d$ interactions). It is a very common feature of LDA calculations, that the gap 
between valence bands and ligand bands is too small. Comparing calculated band structure of 
LaFeAsO with PES experiments, the discrepancy is about 1\,eV.\cite{malaeb1} In one-shot 
LDA+DMFT calculations this gap can artificially be influenced by manually chosen double-counting 
corrections, which is not the case for full self-consistent calculations (see below, 
Sect.~\ref{sect:appendix}). The correction of this gap would only be possible by the explicit 
inclusion of $p$-$d$ interactions, giving also a repulsion between iron and arsenic ions. 
However, these interactions, without further approximations, go well beyond single-site DMFT 
calculations as used here.

Having established the improved description of the crystal structure, we move on to magnetic 
properties of LaFeAsO. 
From now on, we always use the AMF double counting correction,
  ment to be more appropriate for metallic systems. 
We use here again the experimental value for the As $z$ position in 
order to compare more directly to experiments. In \fig{fig:magn} 
we plot the local susceptibility as function of imaginary time for different temperatures. 
Integrating over imaginary time gives the static susceptibility, which is plotted in the lower 
panel of \fig{fig:magn} as black line (open symbols). We see only a
very weak dependence on temperature, consistent with  
experiments.\cite{kamihara1,nomura2008,singh1} Also, the value of $\chi(T)$ is substantially
enhanced compared to free electrons in agreement with
Ref.~\onlinecite{hansmann2010}, a situation often called 'enhanced'  
Pauli-magnetism. Please note that the upturn at the magnetic phase transition 
in experiments is missed here, since we do here calculations only in the paramagnetic phase.

\begin{figure}[t]
  \centering
  \includegraphics[width=0.9\columnwidth]{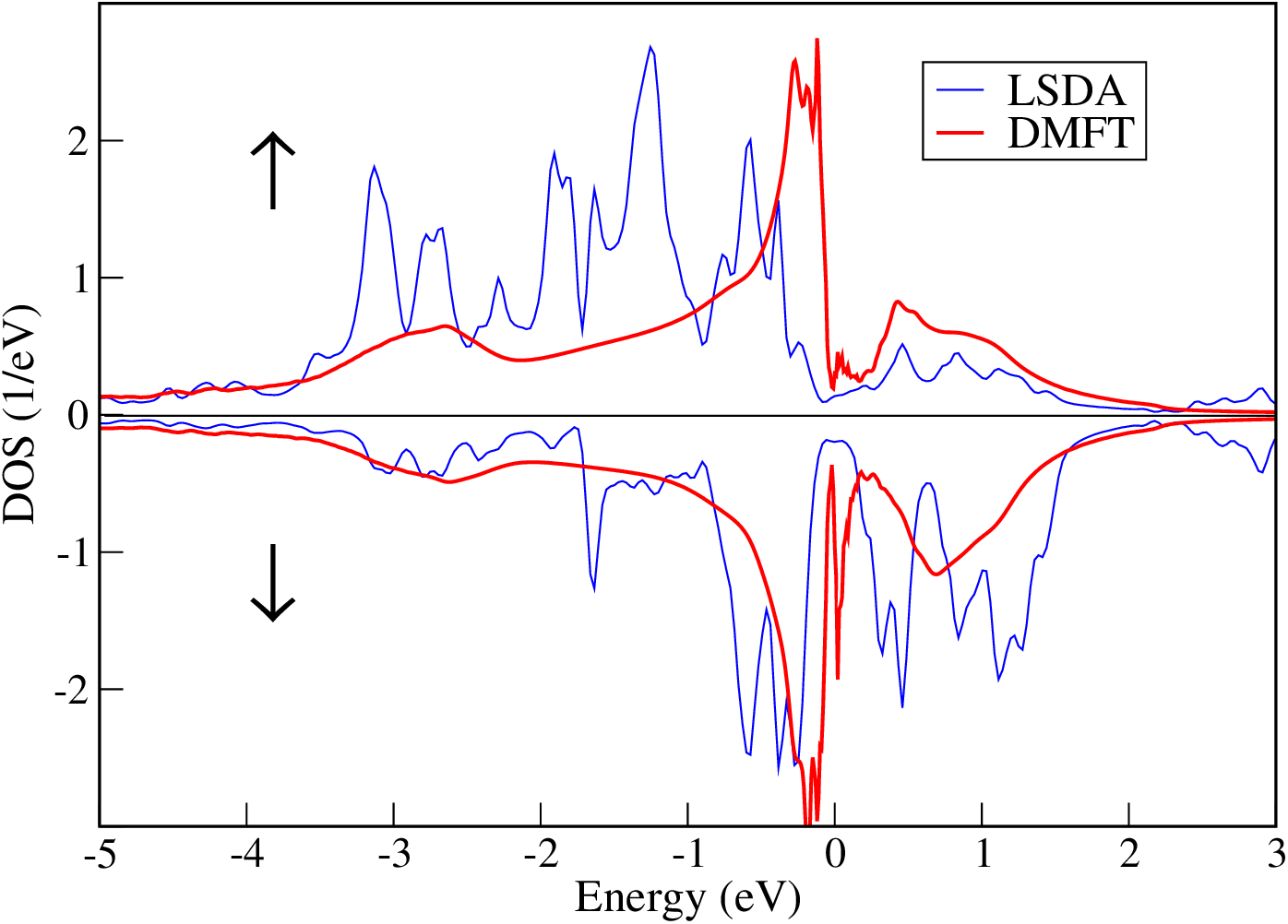}
  \caption{\label{fig:DOS}%
    (Color online) Spin dependent local density of states in the AF
    phase. Blue thin lines: LSDA. Red thicker lines:
    LDA+DMFT 
    using AMF double counting correction. 
  }
\end{figure}

The instantaneous magnetic moment (equal-time correlation function) is
large, roughly 1.95\,\mub. 
However, when looking at the ordered moment at low-temperatures, the situation is different. To 
study the {\it ordered} moment at low temperatures, we performed LDA+DMFT calculations allowing 
for spin polarization at $T=116$\,K, which is well below the magnetic transition temperature of 
$T_N\approx 140$\,K. As magnetic order pattern we assumed the stripe SDW pattern, as 
suggested by LSDA calculations as well as experiments. In order to keep calculations as feasible 
as possible, we used ferromagnetic instead of antiferromagnetic stacking in $c$-direction, but 
since the distance of the iron layers is very large, this
approximation is well justified. We used 
the orthorhombic low-temperature unit cell as given in Ref.~\onlinecite{delacruz1}, with ferromagnetic 
chains running along the short bonds in the $xy$ plane. Doing so, we find an ordered moment of 
$m=0.58$\,\mub, significantly smaller than our LSDA value of
$m=1.74$\,\mub. The value of the magnetic ordered moment is almost
converged in temperature, since calculations for $T=77$\,K give only
slightly larger moments of $m=0.60$\,\mub. 
In a recent comprehensive
LDA+DMFT study a value of $m=0.8$\,\mub was reported for LaFeAsO,\cite{yin2011u} the
difference in the two results coming from the larger interaction
values $U=5.0$ and $J=0.7$ used in Ref.~\onlinecite{yin2011u} (an
estimate on the variation of the magnetic moment as function of
parameters has been given in Ref.~\onlinecite{yin2011}). For
comparison, in the first LDA+DMFT study of the ordered magnetic
moment,\cite{yin2011} done for BaFe$_2$As$_2$, a similar reduction of the magnetic moment to
$m\approx 0.9$\,\mub{} has been found. 
The temporal fluctuations, which are very strong in the LaFeAsO compound due to its quite itinerant 
nature, hinder the instantaneous moments from complete ordering, leaving only a fraction of the 
moment in the ordered state. Similar arguments have been given for the reduction of the moment 
in Ref.~\onlinecite{yin2011}. 

Our findings are in qualitative
agreement with a recent study on the quenching of the magnetic
moment.\cite{hansmann2010} However, the former study has been done in the
{\em paramagnetic} phase, focusing on the influence of local quantum
fluctuations on the local moment. A direct comparison of the values of
magnetic moments is therefore not appropriate. 

The reduction of the ordered moment can also be seen in the local
density of states as shown in \fig{fig:DOS}, where we plot the momentum
integrated spectral function for the Fe 3d electrons. Real frequency
data has been obtained by using the stochastic Maximum Entropy method.\cite{beach_ME}
In LSDA the splitting between
majority and minority spins is large, whereas we see only a small gap
in the LDA+DMFT spectra due to the smaller moment.

\section{Conclusions}\label{sect:conclusions}

In this paper, we presented an extension of the previously introduced
LDA+DMFT approach\cite{aichhorn2009} based on the augmented linearized
plane wave basis to full charge self-consistency, and applied this 
approach to structural and magnetic properties of the iron
superconductor LaFeAsO. We calculated the total energy as function of
the pnictogen height, and found that the inclusion of correlation
effects shift the minimum position from $z=0.632$ to roughly
$z=0.644$, a much better agreement with the experimental value of
$z=0.651$. This increased distance of the As ion from the Fe plane is
due to the high spin state of iron, which is formed due to local
interactions. 

Considering the magnetic properties, we calculated the local spin
susceptibility and found that it shows very weak temperature
dependence in the paramagnetic state, in accordance with enhanced
Pauli magnetism. 
In the low-$T$
SDW phase, we calculated the ordered moment in the stripe-like
antiferromagnetic phase, and found a moment of $m\approx 0.6$\,\mub,
in much better agreement with experimental values than the LSDA value,
which can be (for the experimental crystal structure) as high as
2\mub. 

In summary, the inclusion of correlation effects significantly
improve {\em both structural and magnetic} properties of LaFeAsO within one
set of parameters. This strongly points to the importance of local
quantum fluctuations and correlations for the physics of iron-based
superconductors. 

\acknowledgments

We acknowledge fruitful discussions with J. Mravlje, S. Biermann, 
V. Vildosola, M. Ferrero, and O. Parcollet.
This work was supported by IDRIS/GENCI (project 101393).
M.A. gratefully acknowledges financial support from the Austrian
Science Fund (projects J2760, F4103, and P18551).
L.P. acknowledges financial support of Link\"oping Linnaeus Initiative
for Novel Functional Materials(LiLi-NFM) and  Swedish Research
Council(VR) as well as computational resources provided by the Swedish
National Infrastructure for Computing (SNIC). A.G. acknowledges the support 
of the Agence Nationale de la Recherche (under grant 'PNICTIDES' ANR 2010 BLAN 
0408 04) and the hospitality of the Universit\'e de Gen\`eve (with support from
the MANEP program).

\appendix

\section{Single-shot versus full charge self-consistency}\label{sect:appendix}

\begin{figure}[t]
  \centering
  \includegraphics[width=0.8\columnwidth]{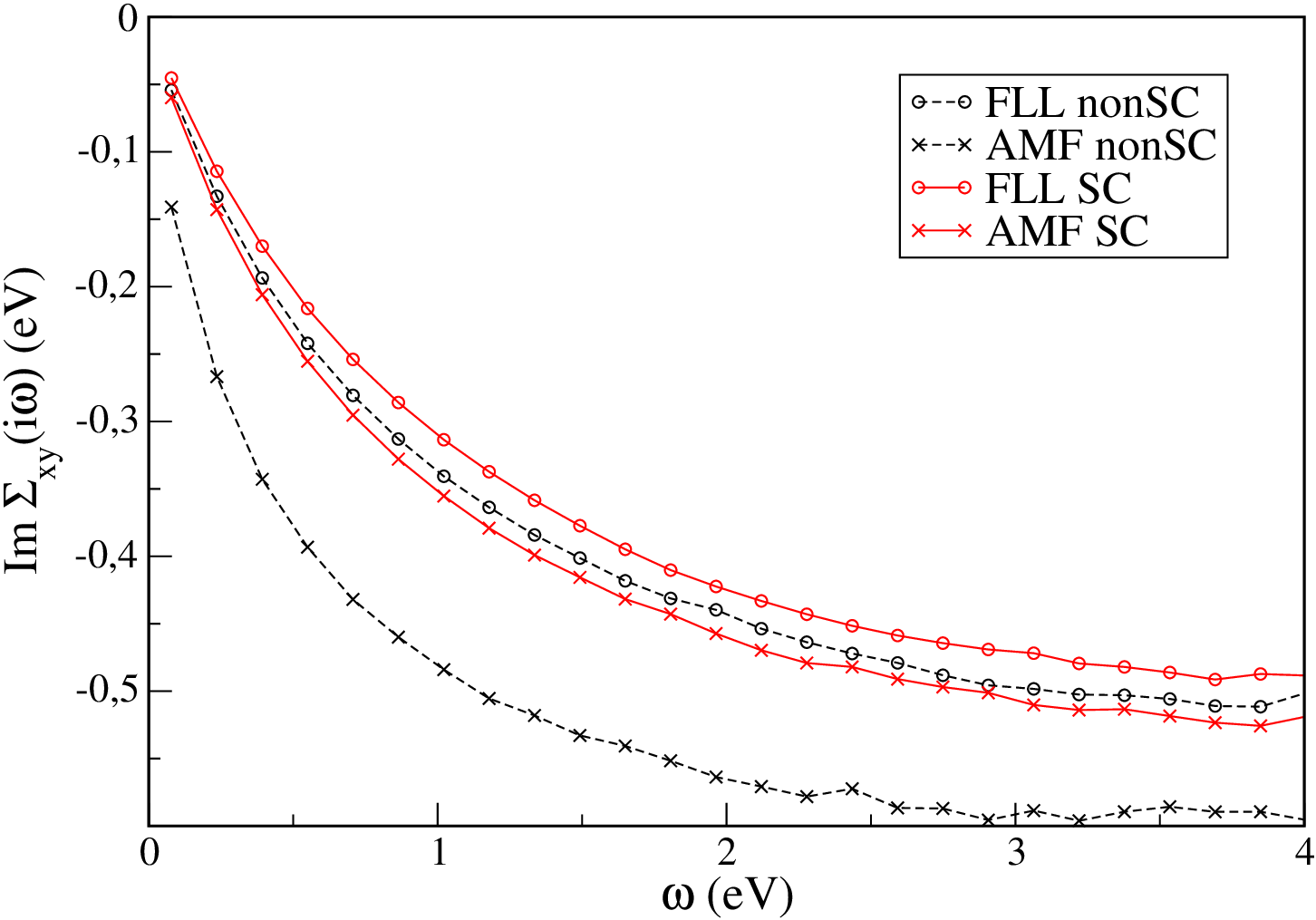}
  \caption{\label{fig:sigma}%
    (Color online) Comparison of the self energies for the $d_{xy}$ orbital for low Matsubara 
    frequencies at $\beta=40$\,eV$^{-1}$. Dashed lines: one-shot non-self-consistent calculations. 
    Solid lines: self-consistent calculations. Circles: FLL double counting. Crosses: AMF double 
    counting.
  }
\end{figure}

In this appendix we show that self-consistent calculations improve over one-shot calculations 
regarding the choice of the double-counting correction. The most straight forward quantity to look 
at is the impurity self-energy on the Matsubara axis, which is not affected by any analytic 
continuation problem to real frequencies. \fig{fig:sigma} shows the result for the iron $d_{xy}$ 
orbital. It is obvious that non-self consistent calculations give a sizable discrepancy between 
FLL and AMF double counting, which is largely canceled in self-consistent calculations. From this 
plot we can also see, that the FLL one-shot calculation is in better agreement with the self-consistent 
calculations, whereas AMF is far off.

\begin{figure}[t]
  \centering
  \includegraphics[width=0.8\columnwidth]{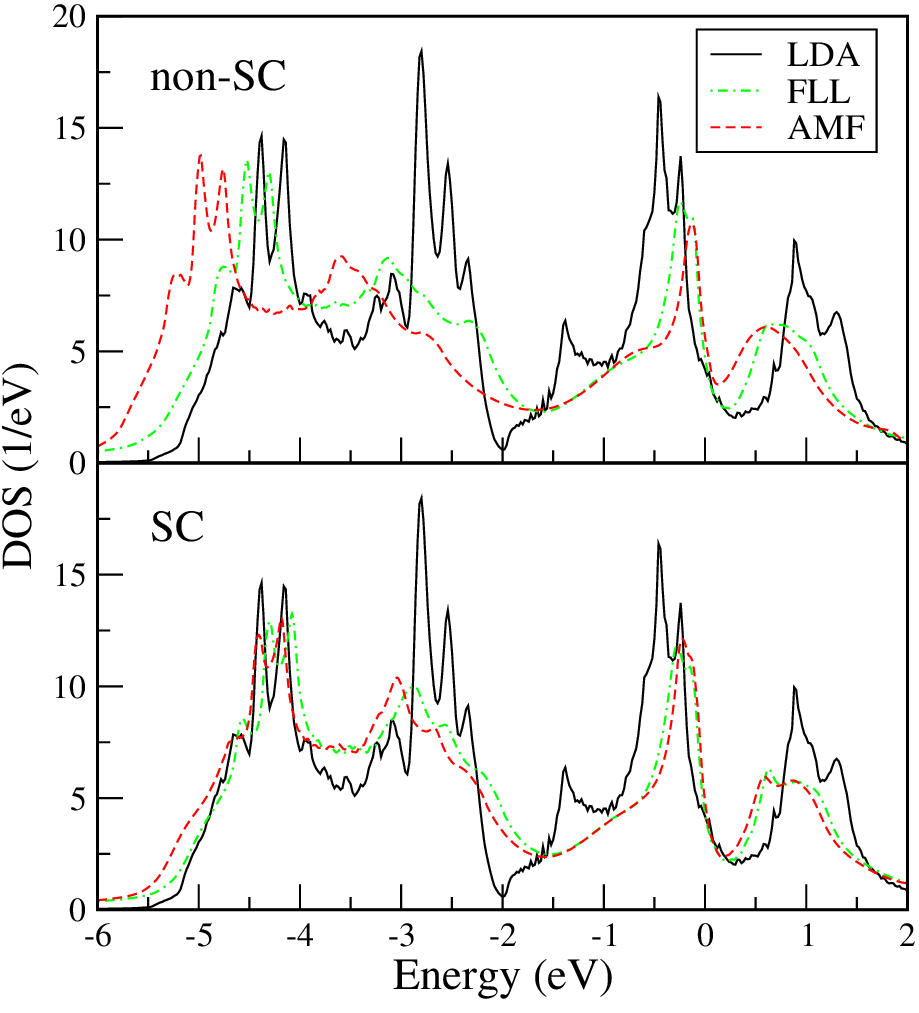}
  \caption{\label{fig:spectra}%
    (Color online) Comparison of the k-summed spectral function for different LDA+DMFT calculations. 
    Black solid lines: LDA result. Green dot-dashed: LDA+DMFT, FLL. Red dashed: LDA+DMFT, AMF. Upper panel: 
    non-self-consistent. Lower panel: self-consistent calculations. 
  }
\end{figure}

Going to the real axis, we can look at the momentum integrated spectral function, and compare it 
with its LDA result, shown in \fig{fig:spectra}. Again, similar as discussed above, the agreement 
between different calculations is much better in the self-consistent case (lower panel), and one-shot 
FLL is again in better agreement. A striking difference between non-self-consistent and self-consistent 
calculations is that there is no spurious shift of the As and O $p$ states due to the approximate 
nature of the double counting correction. Both calculations show the features largely related to As 
and O at basically the same energy as in the LDA calculation. As a result, it is not easily possible 
in self-consistent calculations to use a manually adjusted double counting correction for increasing 
or decreasing the $p$-$d$ gap.

\section{LDA+DMFT charge density within the (L)APW basis set}\label{sect:appendix_chd}

\subsubsection{Charge density within MT-spheres}\label{subsec_MT_dens}

In the LAPW framework the basis functions within an MT-sphere $\alpha$
are the radial solutions $u_{l}^{\alpha,\sigma}$ of the
Schr\"{o}dinger equation  labeled by the orbital quantum number $l$
and spin $\sigma$ and evaluated at a certain linearization energy
$E^{\alpha}_{l1}$ and their corresponding energy derivatives
$\dot{u}_{l}^{\alpha,\sigma}$ evaluated at the same energy. Additional
radial solutions can be  introduced to account for semicore states,
they are evaluated at a corresponding energy  $E^{\alpha}_{l2}$ of the
semicore band. The angular and spin dependence for the solutions
within the MT-spheres is given by corresponding spherical harmonics
$Y^l_m(\hat{r})$ and spinors $\chi_{\sigma}$. The functions
$u_{l}^{\alpha,\sigma}(E^{\alpha}_{l1}) Y^l_m \chi_{\sigma}$,
$\dot{u}_{l}^{\alpha,\sigma}(E^{\alpha}_{1l}  Y^l_m \chi_{\sigma}$
for valence and $u_{l}^{\alpha,\sigma}(E^{\alpha}_{2l})  Y^l_m
\chi_{\sigma}$ for semicore  states will contribute to a given
eigenvector $\Psi_{k\nu}$ with the corresponding coefficients
$A_{lm}^{\nu,\alpha}(\mathbf{k},
\sigma)$,$B_{lm}^{\nu,\alpha}(\mathbf{k},\sigma)$  and
$C_{lm}^{\nu,\alpha}(\mathbf{k},\sigma)$, respectively, as defined in
Ref.~\onlinecite{aichhorn2009}.  

Let us designate the set of these basis functions
$\{u_{l}^{\alpha,\sigma}(E^{\alpha}_{1l}) Y^l_m \chi_{\sigma},
\dot{u}_{l}^{\alpha,\sigma}(E^{\alpha}_{1l}) Y^l_m
\chi_{\sigma},u_{l}^{\alpha,\sigma}(E^{\alpha}_{2l}) Y^{l}_{m}
\chi_{\sigma} \}$ for a given MT sphere $\alpha$ and quantum numbers
$l$, $m$, $\sigma$ as $x^{\alpha,\sigma}_{l}Y^l_m \chi_{\sigma}$ and
the set of corresponding coefficients
$\{A_{lm}^{\nu,\alpha}(\mathbf{k},\sigma),B_{lm}^{\nu,\alpha}(\mathbf{k},\sigma),
C_{lm}^{\nu,\alpha}(\mathbf{k},\sigma)  
\}$ with which they contributes to a given eigenvector $\Psi_{k\nu}$
as $S_{lm}^{\nu,\alpha}(\mathbf{k},\sigma)$. Hence, within a given
MT-sphere the KS eigenvector $\Psi_{k\nu}^{\sigma}(\br)$ is expanded
as $\sum_{l m i} S_{lmi}^{\nu,\alpha}
(\mathbf{k},\sigma)x^{\alpha,\sigma}_{li}(r) Y^l_m(\hat{r})
\chi_{\sigma}$, where $i$ runs over all radial functions $\{x\}$ and
corresponding coefficients $\{S\}$. 

Using those designations, the charge density contribution from the
states within the energy window $\mathcal{W}$ (the second term in RHS
of Eq.~\ref{rho_general_dmft}) can be rewritten for a given MT-sphere
$\alpha$ and for a given spin as 
\begin{align}
\rho^{{W}}_{\sigma \alpha}(\mathbf{r}) & =\sum_{k}\sum_{l
  l'}\sum_{ij}x^{\alpha\sigma}_{li}(r)x^{\alpha\sigma}_{l'j}(r) \nonumber\\  
  &\times \sum_{mm'}   Y^{l}_m(\hat{r})\left(Y^{l'}_{m'}(\hat{r}\right)^* \nonumber\\
  &\times \sum_{\nu\nu' \in \mathcal{W}}S_{lmi}^{\nu,\alpha}(\mathbf{k},\sigma) \left(S_{l'm'j}^{\nu',\alpha}(\mathbf{k},\sigma)\right)^* N^{\bk}_{\nu\nu'}. \label{rho_MT_full}
\end{align}
To represent the angular dependence of the charge density it is
expanded, within a given MT-spheres, in real spherical harmonics
$Y^{Rl}_m(\hat{r})$, 
\begin{equation}\label{rho_MT}
\rho^{\mathcal{W}}_{\sigma \alpha}(\mathbf{r})=\sum_{lm} \rho_{\sigma \alpha}^{lm}(r) Y^{Rl}_m(\hat{r}),
\end{equation} 
For $\rho_{\sigma \alpha}^{lm}(r) = \int d \hat{r}
\rho^{\mathcal{W}}_{\sigma \alpha}(\mathbf{r}) Y^{Rl}_m(\hat{r})$ one
obtains from (\ref{rho_MT_full})

\begin{equation}\label{rho_MT_lm}
\begin{split}
\rho^{l_1 m_1}_{\sigma \alpha}(r) & =\sum_{k}\sum_{l l'}\sum_{ij}x^{\alpha\sigma}_{li}(r)x^{\alpha\sigma}_{l'j}(r)  \\ 
 & \times\sum_{mm'}C^{l'm'}_{lm l_1 m_1}  \\
 & \times\sum_{\nu\nu' \in
   \mathcal{W}}S_{lmi}^{\nu,\alpha}(\mathbf{k},\sigma)
 \left(S_{l'm'j}^{\nu',\alpha}(\mathbf{k},\sigma)\right)^*
 N^{\bk}_{\nu\nu'} , 
\end{split}
\end{equation}
where $C^{l'm'}_{lm l_1 m_1} = \int d\Omega Y^{l}_m(\Omega)\left(Y^{l'}_{m'}(\Omega)\right)^*Y^{Rl_1}_{m_1}(\Omega)$ are the corresponding Gaunt coefficients.

\subsubsection{Charge density in the interstitial region}\label{subsec_inter_dens}

In the interstitial region  the LAPW basis functions are plain waves
$\frac{1}{\sqrt{V}}e^{i(\mathbf{k}+\mathbf{G})\mathbf{r}}$, where $G$
is the reciprocal lattice vector and V is the unit cell volume,
contributing to a given KS eigenvector $|\Psi^{\sigma}_{k\nu}\rangle$
with the corresponding coefficients
$a_{G}^{\nu}(\bk,\sigma)$. Substituting this into
Eq.~\ref{rho_general_dmft} one obtains the contribution of the states
within the energy window $\mathcal{W}$  to the charge density  in the
interstitial: 
\begin{widetext}
\begin{equation}\label{rho_inter}
\begin{split}
\rho^{\mathcal{W}}_I(\mathbf{r}) &= \frac{1}{V} \sum_{\bk} \sum_{\nu\nu' \in \mathcal{W} }  
\left[\sum_{G} a_{G}^{\nu}(\bk,\sigma)e^{i\mathbf{k_G r}} \sum_{G'} \left(a_{G'}^{\nu'}(\bk,\sigma)\right)^*e^{-i\mathbf{k_{G'} r}}\right] N^{\bk}_{\nu\nu'} \\
&= \frac{1}{V} \sum_{\bk} \sum_{\nu\nu' \in \mathcal{W} }  
\left[\sum_{G} a_{G}^{\nu}(\bk,\sigma)e^{i\mathbf{G r}} \sum_{G'} \left(a_{G'}^{\nu'}(\bk,\sigma)\right)^* e^{-i\mathbf{G'r}}\right] N^{\bk}_{\nu\nu'}, \\
\end{split}
\end{equation}
\end{widetext}
where $\mathbf{k_G}=\mathbf{k}+\mathbf{G}$. 

In the actual computation of (\ref{rho_inter}) one may transform the
interstitial wave function $\sum_{G}
a_{G}^{\nu}(\bk,\sigma)e^{i\mathbf{G r}}$ to a auxiliary mesh in the
real space via the fast Fourier transform.:

\begin{equation}\label{fft}
b^{\nu}_{\mathbf{R}}(\bk,\sigma)=\sum_{G}a_{G}^{\nu}(\bk,\sigma)e^{i\mathbf{G R}},
\end{equation}
therefore getting rid off the double sum over $\mathbf{G}$ and
$\mathbf{G'}$ in (\ref{rho_inter}). The charge density on the
axillary mesh then reads: 
\begin{equation}\label{rho_real}
\rho^I(\mathbf{R})=\frac{1}{V} \sum_{\bk} \sum_{\nu\nu' \in
  \mathcal{W} }
b^{\nu}_{\mathbf{R}}(\bk,\sigma)(b^{\nu'}_{\mathbf{R}}(\bk,\sigma))^*
N^{\bk}_{\nu\nu'}, 
\end{equation}
which is then again transformed back to the reciprocal space via
inverse FFT: 
\begin{equation}\label{rho_recip}
\rho^I(\mathbf{G})=\sum_{R}\rho^I_{\mathbf{R}}e^{-i\mathbf{G R}}.
\end{equation}

\end{document}